\documentclass[pra,twocolumn]{revtex4}

\usepackage{graphicx}
\usepackage{dcolumn}
\usepackage{bm}

\newcommand{\bo}{\raise-1mm\hbox{\Large$\Box$}}

\newcommand{\rr}{\mathbf{r}}
\newcommand{\kk}{\mathbf{k}}
\usepackage{epsfig}
\usepackage{latexsym}
\usepackage{mathrsfs}
\usepackage{setspace}
\usepackage{amsmath}
\usepackage{floatflt}
\usepackage{wrapfig}
\long\def\symbolfootnote[#1]#2{\begingroup
\def\thefootnote{\fnsymbol{footnote}}\footnote[#1]{#2}\endgroup}

\begin{document}

\title{Angular collapse of dipolar Bose-Einstein condensates}

\author{Ryan M. Wilson\email{rmw@colorado.edu}}
\affiliation{JILA, NIST and Department of Physics, University of Colorado, Boulder, Colorado 80309-0440, USA}

\author{Shai Ronen}
\affiliation{JILA, NIST and Department of Physics, University of Colorado, Boulder, Colorado 80309-0440, USA}

\author{John L. Bohn}
\affiliation{JILA, NIST and Department of Physics, University of Colorado, Boulder, Colorado 80309-0440, USA}

\date{\today}

\begin{abstract}
We explore the structure and dynamics of dipolar Bose-Einstein condensates (DBECs) near their threshold for instability.  Near this threshold a DBEC may exhibit nontrivial, biconcave density distributions, which are associated with instability against collapse into ``angular roton'' modes.  Here we discuss experimental signatures of these novel features.  In the first, we infer local collapse of the DBEC  from the experimental stability diagram.  In the second, we explore the dynamics of collapse and find that a nontrivial angular distribution is a signature of the DBEC possessing a biconcave structure.
\end{abstract}

\maketitle

\section{Introduction}
\label{sec:intro}

The role of dipole-dipole interactions in the structure and dynamics of quantum many-body systems is at the forefront of both theoretical and experimental research.  The dipole-dipole interaction is long-ranged and anisotropic, in contrast to the isotropic $s$-wave contact interaction, and leads to novel physics in ultracold spinor~\cite{Vengalattore08,Kawaguchi07}, fermonic~\cite{Baranov08PR,Baranov02PRA,Baranov04PRL,KKScience08,Yi04} and bosonic~\cite{Santos03a,Ronen07,Ronen06a,Wilson08,Saito09PRL} many-body systems.

For example, the anisotropy of the dipole-dipole interaction leads to unusual properties of a Bose-Einstein condensate (BEC).  The condensate will experience ``magnetostriction,'' wherein its aspect ratio does not match that of the trap in which it is confined~\cite{Lahaye07}.  Additionally, a dipolar BEC (DBEC) is expected to exhibit novel density profiles, for instance a biconcave shape where the maximum density occurs at the periphery of the condensate rather than at its center~\cite{Ronen07}, or else more elaborate densities for non-cylindrically symmetric traps~\cite{Dutta07}.  In contrast to magnetostriction, these density profiles have not thus far been directly imaged in experiments since shapes in the dentiy profiles likely wash out upon expansion of the condensate.

Unusual density distributions are, however, intimately related to a more clearly observable property of the gas, namely, its macroscopic collapse.  Because the dipole-dipole interaction is always attractive for dipoles in a head-to-tail orientation, a DBEC will always become unstable for a sufficiently large dipole density and collapse inward, just as is the case for atomic BECs with attractive contact interactions.   The novelty is that, since the dipole-dipole interaction is anisotropic, the density at which the instability occurs depends on the aspect ratio of the harmonic trapping potential that confines the dipoles.

This instability has been probed experimentally in a DBEC of atomic $^{52}\mathrm{Cr}$~\cite{KochNature08a}.  In this experiment the condensate was stabilized by tuning the $s$-wave scattering length to a sufficiently large value.  Instability was then triggered by reducing this scattering length below a critical value.  Strikingly, the observed anisotropic density distribtion of the collapsed cloud was well-reproduced within the mean-field theory~\cite{Lahaye08PRL}.

On the theory side, the detailed mechanism of the collapse has been discussed in~\cite{Parker09,Bohn09}.  The picture that emerges is that for sufficiently oblate traps the collapse of the DBEC occurs via local density fluctuations, rather than a global collapse to the trap's center.  In this case the instability is driven by fluctuations into a soft ``roton''-like mode, which is unique to condensates with dipolar interactions~\cite{Santos03a}.  Depending on the trap's aspect ratio, the roton can have nodal surfaces in either the radial or angular coordinate, for a cylindrically symmetric condensate, and these nodal patterns should dictate the details of the condensate's collapse.  In particular, for condensates with a biconcave density profile, the roton decay mode should always have an angular nodal pattern.  Measuring an angular collapse would then be an experimental signature of the biconcavity.

Thus far there is no direct experimental evidence for the roton or for the local collapse in a DBEC, although we have argued that the data in Ref.~\cite{KochNature08a} supports the idea of a local collapse~\cite{Bohn09}.  Further, Ref.~\cite{Parker09} has explored circumstances of local collapse in various trap geometries, contrasting approches where the collapse is initiated by either rapid or else adiabatic changes in the scattering length.  

In this article we therefore tackle head-on the prospects for observing local collapse.  After some preliminary remarks in Sec.~\ref{sec:prelim}, we proceed in Sec.~\ref{sec:local1} to analyze and extend the experimental result in Ref.~\cite{KochNature08a}.  We argue that for oblate traps the scattering length required to stabilize the condensate can be explained within mean-field theory, but only if the collapse occurs locally.  Further, the distinction between local and global collapse becomes more clear if the number of dipoles is increased.  In Sec.~\ref{sec:local2} we develop an understanding of a more direct measurement of condensate collapse, following the experimental procedure of Ref.~\cite{Lahaye08PRL}, which includes an expansion that allows for imaging of the cloud.  We show that angular structure in the expanded image is a direct signature of biconcave structure.

\section{Preliminaries}
\label{sec:prelim}

Consider a dilute gas of dipolar entities each with dipole moment $d$ polarized in the axial ($z$) direction.  The interaction between two such entities interacting by both the dipole-dipole and contact interactions is given, in cgs units, by
\begin{equation}
\label{V}
V(\rr-\rr^\prime) = d^2 \frac{1-3\cos^2{\theta}}{|\rr-\rr^\prime|^3} + \frac{4\pi \hbar^2 a_s}{M}\delta(\rr-\rr^\prime)
\end{equation}
where $|\rr-\rr^\prime|$ is the distance between the particles, $\theta$ is the angle between $\rr-\rr^\prime$ and the $z$-axis, $M$ is the particle mass and $a_s$ is the $s$-wave scattering length of the particles.  The contact interaction (second term in Eq.~(\ref{V})) is either repulsive ($a_s>0$) or attractive ($a_s<0$), regardless of the orientation of the particles.  The dipole-dipole interaction (first term in Eq.~(\ref{V})), however, changes sign depending on the particle's orientation.  Two dipoles aligned in the direction of their polarization ($\theta=0$) attract each other while two dipoles aligned orthogonal to this direction ($\theta=\pi/2$) repel each other.

We consider such a gas confined by a harmonic potential of the form $U(\rr) = \frac{1}{2} M \omega_\rho^2(\rho^2 + \lambda^2 z^2)$ where $\lambda = \omega_z / \omega_\rho$ is the trap aspect ratio, describing to what degree the trap is prolate ($\lambda<1$) or oblate ($\lambda>1$).  The trapping potential introduces a zero-point contribution to the condensate energy which serves to stabilize the system.  A gas without dipoles but possessing a small negative scattering length proves stable for sufficiently low density at any trap aspect ratio.  The negative scattering length at which the condensate goes unstable scales only weakly with trap aspect ratio.  When the stability threshold is crossed, e.g. when the scattering length becomes sufficiently negative to destabilize the BEC, the condensate undergoes macroscopic collapse.  For purely contact interactions, the mean-field potential of the condensate is directly proportional to the density of the condensate, so collapse occurs where the particle density is greatest, at the center of the trap~\cite{Duine01}.  

By contrast, the trap aspect ratio plays a decisive role in determining the stability of a DBEC.  In a prolate trap, a DBEC behaves much like a BEC with attractive contact interactions.  This geometry favors  attraction between dipoles and will induce a global collapse to the center for a critical dipole-dipole interaction strength.  Vice versa, a DBEC in an oblate trap might be expected to behave much like a BEC with repulsive contact interactions since the dipolar entities are predominately repulsive in this geometry.  However, as shown in~\cite{Ronen07}, there exists a finite critical dipole-dipole interaction strength, for any aspect ratio, at which a DBEC becomes unstable.  The mechanism for collapse in this large $\lambda$ regime, however, is very different than that of a DBEC in a prolate trap or of a BEC with purely contact interactions.

In an oblate trap, the axial trapping frequency is large, which acts to suppress elongation in the trap center, rendering the global collapse unlikely.   Instead, the dipoles in the condensate are expected to form local density maxima whose spatial widths are on the order of the axial harmonic oscillator length $a_z = \sqrt{\hbar / M\omega_z}$. Each such bunch of dipoles then elongates axially, leading to local collapse.  These local density maxima are related to the softening of a roton mode, whose characteristic wavelength $a_z$ sets the scale of the local collapse~\cite{Santos03a,Wilson08}.

\section{Local collapse: evidence from the stability diagram}
\label{sec:local1}

Thus far only one experiment has explored the stability of a DBEC as a function of the trap aspect ratio~\cite{KochNature08a}.  Rather than tune the dipole moment to a critical value, the experiment instead artificially stabilized the condensate by introducing a positive s-wave scattering length via a Fano-Feshbach resonance.  Upon reducing the scattering length below a critical value $a_\mathrm{crit}$, the experiment was able to trigger collapse in the DBEC.

\begin{figure}
\includegraphics[width=\columnwidth]{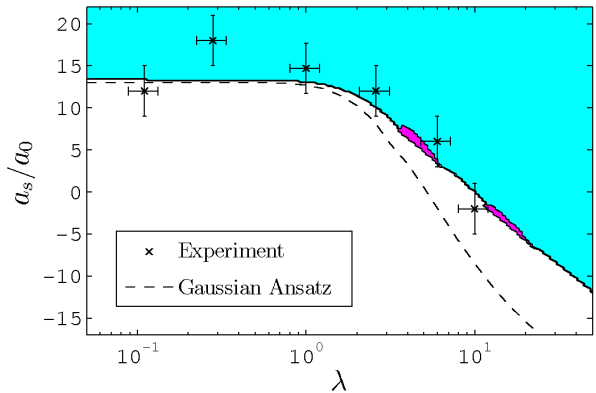}
\caption{\label{fig:N20000}  (color online).  The stability diagram of a DBEC of $N=2\times 10^4$ $^{52}\mathrm{Cr}$ atoms plotted as critical scattering length versus trap aspect ratio.  The points show the experimental results of~\cite{KochNature08a}, the shaded regions show the results of solving the GPE exactly and the dashed line shows the results of the Gaussian \emph{ansatz}.  The theoretical methods disagree as trap aspect ratio $\lambda$ increases, and the exact results fit the experimental data with great accuracy.  The pink (darker) regions are where biconcave structure is predicted on the condensate profile.}
\end{figure}

The resulting experimental stability diagram (reproduced from~\cite{KochNature08a}) is presented in Figure~\ref{fig:N20000} as a plot of the critical scattering length $a_\mathrm{crit}$ versus aspect ratio $\lambda$.  These results represent the measurement performed on a condensate of $N = 2 \times 10^4$ $^{52}$Cr atoms.  For prolate traps, a comparatively large scattering length is required to achieve stability.  As $\lambda$ is increased, the zero-point energy in the axial direction stabilizes the DBEC, and stable condensates are possible with a smaller critical scattering length.

This figure also shows the results of two alternative numerical calculations of critical scattering length.  In one, the theoretical division between stable (shaded) and unstable (unshaded) regions of parameter space is determined by numerically solving the nonlocal Gross-Pitaevskii equation (GPE) using the potential in~(\ref{V}).  A second approach, already employed as an approximation in the experimental paper, shows the dividing line between the stable and unstable regions as a dashed line.  This approximation posits a Gaussian ansatz wave function (normalized to unity):
\begin{equation}
\label{gausspsi}
\psi(\rho,z) = \left(\frac{1}{\pi^{3/2} \sigma_\rho^2 \sigma_z \bar{a}_\mathrm{ho}^3} \right)^2 \exp{\left[ \frac{-1}{2\bar{a}_\mathrm{ho}^2}\left( \frac{\rho^2}{\sigma_\rho^2} + \frac{z^2}{\sigma_z^2} \right)\right]}
\end{equation}
where $\sigma_\rho$ and $\sigma_z$ are the variational parameters and $\bar{a}_\mathrm{ho} = \sqrt{\hbar / M \bar{\omega}}$, where $\bar{\omega} = \sqrt[3]{\omega_\rho^2 \omega_z}$ is the geometric mean trap frequency.  Using this ansatz, the Gross-Pitaevskii energy functional~\cite{BEC2003},
\begin{eqnarray}
\label{GPEf}
E[\psi,\psi^\star] = \int \left[ \frac{\hbar^2}{2M}|\nabla \psi(\rr)|^2 + U(\rr)|\psi(\rr)|^2 \right. \nonumber \\
\left.+ \frac{N-1}{2} |\psi(\rr)|^2 \int \psi^\star(\rr^\prime) V(\rr - \rr^\prime) \psi(\rr^\prime)d\rr^\prime \right] d\rr,
\end{eqnarray}
where $N$ is the condensate particle number, is calculated for a given $\bar{a}_\mathrm{ho}$ to determine whether the energy $E[\psi,\psi^\star]$ has a minimum, and thus to determine if the condensate is energetically stable.  The presence of a  minimum, local or global, corresponds to the presence of a stable ground state.  A key feature of the Gaussian trial wave function is that it always places the maximum density in the condensate's center, i.e., it is incapable of describing local collapse.  For prolate traps, the maximum density {\it is} in the center.  In this case the Gaussian \emph{ansatz} and the numerical solution to the GPE agree with each other on the critical scattering length, and they both are in good agreement with the experimental result.

Care must be taken, however, using this approximation for oblate condensates.  This can be seen in the $\lambda>1$ region of the stability diagram in Figure~\ref{fig:N20000}, where the Gaussian \emph{ansatz} predicts a lower critical scattering length than does the GPE.  We attribute this difference to the ability of the GPE to model local collapse.  Indeed, for larger aspect ratios we observe local collapse into roton-like modes, as we will discuss in the next section, and as have been reported in Ref. \cite{Parker09}.  Further, the experimental determination of $a_\mathrm{crit}$ tends to show better agreement with the GPE prediction than with that of the Gaussian \emph{ansatz}.  We interpret this as experimental support for the occurence of local collapse, albeit somewhat indirect evidence.  However, the roton modes involved in collapse might have either radial or angular nodal structure.  This experiment does not make this distinction.

\begin{figure}
\includegraphics[width=\columnwidth]{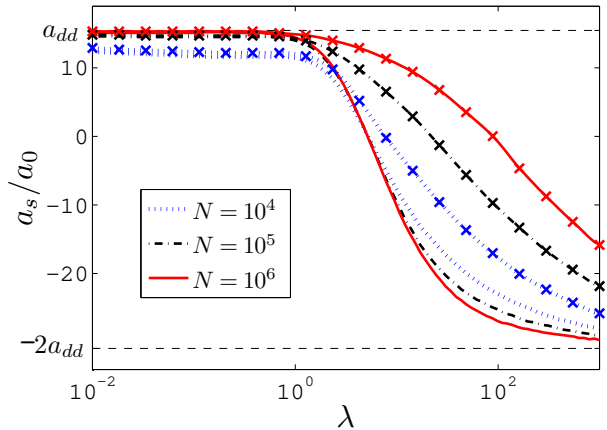}
\caption{\label{fig:stab}  (color online).  For a DBEC of atomic $^{52}\mathrm{Cr}$, this figure illustrates the critical $s$-wave scattering length (below which the DBEC is unstable) as a function of trap aspect ratio $\lambda$ in units of the Bohr radius for $\bar{\omega} = 2\pi\times 700 \, \mathrm{Hz}$.  The blue (dotted), black (dot-dashed) and red (solid) lines correspond to $N=10^4$, $N=10^5$ and $N=10^6$, respectively.  The lines without crossed symbols are the results of the Gaussian \emph{ansatz} and lines with crossed symbols are the results of the exact solution of the GPE.  Notice how, as $N$ is increased, the Gaussian \emph{ansatz} predicts a more stable condensate while the exact solution predicts a less stable condensate.}
\end{figure}

Within the uncertainty in the experiment, the data in Figure~\ref{fig:N20000} discriminates between the two methods, but one may wish for a clearer discrimination.  We therefore consider cases where the atom number is increased.  The critical scattering length $a_\mathrm{crit}$ is shown in Figure~\ref{fig:stab} for DBECs with atom numbers of $N=10^4$, $10^5$, and $10^6$.  For a given trap, increasing the number of dipoles increases the relative importance of the dipole-dipole interaction, which acts to further destabilize the condensate.  Thus, as predicted by the GPE, $a_\mathrm{crit}$ increases with increasing atom number.  Vice versa, the  Gaussian \emph{ansatz} predicts a more stable condensate with increasing atom number.  The difference between the two theoretical approaches could then be clearly distinguished in such an experiment.

Although Figure~\ref{fig:stab} plots only the domain of aspect ratios $10^{-2} < \lambda < 10^3$, it is straightforward to obtain the stability thresholds in the $\lambda \ll 1$ and the $\lambda \gg 1$ limits for fixed $\bar{\omega}$.  These limits are usefully described in terms of the  characteristic dipole length, given by (in CGS units)~\cite{KochNature08a}
\begin{equation}
\label{add}
a_{dd} = \frac{d^2 M}{3 \hbar^2}.
\end{equation}
In the limit of large dipole-dipole interactions, $N a_{dd} / \bar{a}_\mathrm{ho} \gg 1$, the interaction-dependent term in $E[\psi,\psi^\star]$ dominates over the other terms since it scales with $N$.  Thus, in this limit, the condensate is unstable if this interaction-dependent term is negative.  Now, to treat the limits of very small and large $\lambda$, we consider, respectively, the harmonic trapping potential in the limits $\omega_z \rightarrow 0$ and $\omega_\rho \rightarrow 0$.

Taking the limit $\omega_z \rightarrow 0$ corresponds to an infinitely prolate, or cigar-shaped trap.  In this geometry, the dipolar mean-field term reduces to a simple coupling to the condensate density, $d^2 \int \psi^\star(\rho^\prime) \frac{1-3\cos^2{\theta}}{|\rr-\rr^\prime|^3} \psi(\rho^\prime) d\rr^\prime = -4\pi \hbar^2 a_{dd} |\psi(\rho)|^2/ M$ because the dipole-dipole interaction reduces to a delta-function in $\rho$ for this geometry.  Thus, in the quasi-1D geometry, the total mean-field term becomes $4\pi\hbar^2\left(a_s - a_{dd}\right)|\psi(\rho)|^2$.  Similarly for the limit  $\omega_\rho \rightarrow 0$, corresponding to an infinitely oblate trap, the dipolar mean-field term reduces to $ 8\pi \hbar^2 a_{dd} |\psi(z)|^2/ M$, giving a total mean-field term of $4\pi\hbar^2\left( a_s+2a_{dd} \right)|\psi(z)|^2$ for this geometry.  Thus, in the limit $N a_{dd} / \bar{a}_\mathrm{ho} \gg 1$, we find that the DBEC is unstable when $a_s < a_{dd}$ for $\lambda \ll 1$ and the DBEC is unstable when $a_s < -2 a_{dd}$ for $\lambda \gg 1$.  These limits are indicated in Figure~\ref{fig:stab}, and the mean-field calculations are detailed in Appendix~\ref{app:meanfield}.
\section{Local Collapse: evidence from the  collapsed cloud}
\label{sec:local2}

To take a closer look at the nature of collapse, it is necessary to track the collapse itself as a function of time.  This, too, has been achieved in the $^{52}\mathrm{Cr}$ DBEC experiments, which are excellently reproduced by time-dependent mean-field theory~\cite{Lahaye08PRL}.  These experiments have not, however, focused directly on observing consequences of local collapse.  Here we discuss the prospects of making such a measurement.

Briefly, in such an experiment the scattering length is altered from a value where the condensate is stable against collapse to a somewhat lower value $a<a_\mathrm{crit}$.  After this transition, the atoms begin their collapse into high density regions where three-body recombination takes over, ejecting atoms from the trap.  The trap is generally released after some hold time, to expand the cloud for imaging.  The resulting density patterns show intricate shapes and depend on details such as whether the passage from stable to unstable is adiabatic or diabatic~\cite{Parker09}.

\subsection{Modes of instability}

The underlying physics of the instability and collapse is determined by the softening of the roton modes.  We compute these modes by solving the Bogoliubov de Gennes equations as in Ref.~\cite{Ronen06a}, exploiting the cylindrical symmetry of the system.  Namely, we make the quasiparticle \emph{ansatz}, 
\begin{eqnarray}
\psi(\rr,t) \rightarrow \left[\psi(\rho,z) + \delta u(\rho,z)e^{i(m\varphi - \omega t)} + \right. \nonumber \\ 
\left. + \delta v^\star(\rho,z)e^{-i(m\varphi - \omega t)}\right]e^{-i\mu t}
\end{eqnarray}
where $\omega$ is the quasiparticle energy, $m$ is the projection of the quasiparticle momentum onto the $z$-axis, $\mu$ is the chemical potential of the ground state $\psi(\rho,z)$ and $\delta \ll 1$ to ensure that the quasiparticles have small amplitudes.  By solving the BdG equations for various $m$ quantum numbers, we determine whether the DBEC is dynamically stable or unstable by determining whether the quasiparticle energy is purely real or has a nonzero imaginary part, respectively~\cite{BEC2003}.

\begin{figure}
\includegraphics[width=\columnwidth]{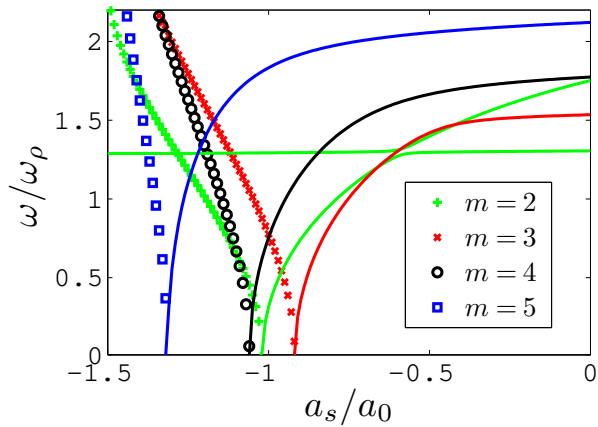}
\caption{\label{fig:ex8} (color online).  The real and imaginary parts of the low-lying BdG modes for a condensate of $N=10^4$ $^{52}\mathrm{Cr}$ atoms with mean trap frequency $\bar{\omega} = 2\pi \times 700$ Hz and trap aspect ratio $\lambda = 8$, plotted as a function of the $s$-wave scattering length $a_s$.  The real parts are represented by solid lines and the imaginary parts, developing where the real parts go to zero, are represened by markers.  The $m=3$ mode, being the first to develop a nonzero imaginary energy, serves to define $a_\mathrm{crit}$ for this aspect ratio.}
\end{figure}

Figure~\ref{fig:ex8} illustrates the mode softening for a DBEC containing $N=10^4$ $^{52}\mathrm{Cr}$ atoms at an aspect ratio $\lambda=8$.  Plotted is the energy of the excitation as a function of the scattering length $a_s$, labeled by its azimuthal angular momentum quantum number $m$.  The solid lines depict the real parts of these energies, while the symbols represent their imaginary parts.  As $a_s$ diminishes, the energies of these modes drop to zero, and thereafter become purely imaginary. The first such transition, at $a_s \sim -0.9 a_0$, identifies the scattering length at which the DBEC is dynamically unstable, since any small perturbation is capable of exciting this mode, which then grows exponentially in time.  Thus an unstable condensate quickly grows high-density peaks in regions defined by the antinodes of these modes.  

Figure~\ref{fig:ex8} is a particular example illustrating the modes that contribute to instability at a particular aspect ratio $\lambda=8$.  At this aspect ratio the condensate's density exhibits a biconcave shape, and so decay into angular rotons is expected.  We stress that at all aspect ratios where the maximum density lies at the center rather than at the periphery, the rotons responsible for instability are always $m=0$ modes that do not exhibit an angular structure.  This connection is essential to connecting observed angular decay circumstantially to biconcave structure.

Regardless of whether the roton is purely radial or angular in nature, it leads a DBEC to instability at a fixed length scale, as mentioned in Sec.~\ref{sec:prelim}, having wavelength $\sim 2\pi a_z$.  As the trap aspect ratio $\lambda$ is increased, the ratio of the axial to the radial harmonic oscillator lengths, $a_z/a_\rho$, is decreased, so more roton wavelengths can fit around the circumference of the condensate for larger $\lambda$.  For biconcave condensates, this results in angular rotons with larger $m$ quantum number being responsible for instability for larger $\lambda$, since more angular nodes can fit into the condensate in this regime.  For $N=10^4$ $^{52}\mathrm{Cr}$ atoms in a trap with $\lambda=8$ and $\bar{\omega} = 2\pi \times 700$ Hz, this mode happens to have $m=3$.  Indeed, the circumference of the region of maximum particle density in this biconcave condensate is $\sim 6\pi a_z$, or three roton wavelengths.

\subsection{Numerics and the ``ideal experiment''}

The mode that brings about the dynamical instability determines not only the scattering length at which the condensate will collapse, but also \emph{how} the condensate will collapse as the stability threshold is crossed.  Consider preparing a DBEC of $N=10^4$ $^{52}\mathrm{Cr}$ atoms just above the stability threshold in a trap with aspect ratio $\lambda=8$.  These are the collective modes whose energies are shown in Figure~\ref{fig:ex8}.  A small jump in scattering length to a value just below $a_\mathrm{crit}$ would cause the condensate to go unstable by a macroscopic occupation of the $m=3$ mode that has a nonzero imaginary energy at this scattering length.  The density of the condensate during the collapse would change, on a time scale $\tau \sim 2\pi / \mathrm{Im}[\omega]$, as the atoms macroscopically occupy three clumps that self-attract in the $z$-direction.  

Decay of the condensate into a roton mode with $m>0$ requires breaking the condensate's initial cylindrical symmetry by introducing fluctuations into the mode.  In an experiment this is caused by thermal fluctuations, but in our calculation we must make this happen artificially. To do this, we seed the condensate wave function by adding to it a small contribution of the excited state mode:
\begin{eqnarray}
\label{one_fluctuation}
\psi({\vec r}) \rightarrow \psi(\rho, z) + 0.01 e^{2 \pi i \alpha}
e^{3i\phi} {\bar u}_3(\rho, z),
\end{eqnarray}
where $e^{3 i\phi}$ describes the basic angular variation of the roton mode, and $\alpha$ is an additional phase that will determine the overall rotation of the collapsed condensate.  In the absence of a seed like this, the numerical solution remains at its unstable equilibrium for a time long compared to the natural lifetime $2 \pi /\mathrm{Im}[\omega]$.  The apparent lifetime in this case is determined by the time before roundoff error starts to affect the time evolution of the GPE.  However, once the wave function is seeded as above, the decay occurs on the expected time scale.

\begin{figure}
\includegraphics[width=\columnwidth]{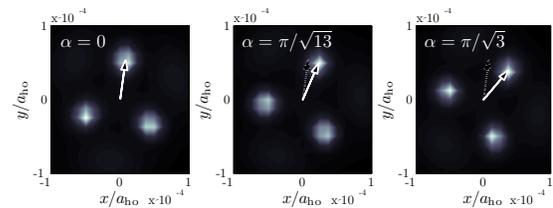}
\caption{\label{fig:phase} Collapsed condensate in a trap with mean frequency $\bar{\omega} = 2\pi \times 700$ Hz and aspect ratio $\lambda = 8$ after $10.5$ ms.  The perturbation for the collapse is controlled to have $m=3$ symmetry and the global phases as shown in the frames.  Each frame corresponds to a different value of the initial phase $\alpha$.  The collapsed condensates are rotated by $\alpha / 3$, ensuring that the finite grid size does not influence the small length scale dynamics of the condensate collapse.}
\end{figure}

After the collapse is triggered, the condensate indeed forms the three clumps as expected, as seen in Figure~\ref{fig:phase}.  Shown is the density of particles, as viewed in the $x$-$y$ plane, i.e., looking down from the axis of the dipoles' polarization.  Each peak was initially seeded by a density fluctuation at the antinode of the $m=3$ excited state roton wave function.  Thus the three peaks are uniformly equally spaced in angle, as befits the symmetery of the mode.  An angular display of this sort would provide unambiguous evidence for nonlocal collapse.  Moreover, the fact that the collapse occured in an angular coordinate provides indirect evidence for the biconcave structure of the initial state.

While the relative positions of the three peaks in this experiment are well-defined by the symmetry of the roton mode, there is still an overall undetermined angle of rotation of the whole pattern.  Numerically, this is set by the angle $\alpha$ in (\ref{one_fluctuation}).  Since the angular dependence of the condensate density, with this wave function, is proportional to $\cos{(3\varphi + \alpha)}$, we expect that, if there is no unphysical dependence on the numerical grid, the collapse will occur rotated by an angle $\alpha/3$ for any initial phase $\alpha$.  Indeed, we find that the collapse dynamics are unaffected by the grid, as is illustrated in Figure~\ref{fig:phase}.  Here, we input the initial phases $\alpha = \pi / \sqrt{13}$ and $\alpha = \pi / \sqrt{3}$ and find that the collapsed wave function is rotated by exactly these phases times $1/3$.  Although not shown here, simulations for other initial phases give the same results.  Based on this ability to reproduce the same angular pattern, but rotated in a predictable way, we conclude that the underlying Cartesian grid is adequate to describe this collapse.

\begin{figure}
\includegraphics[width=\columnwidth]{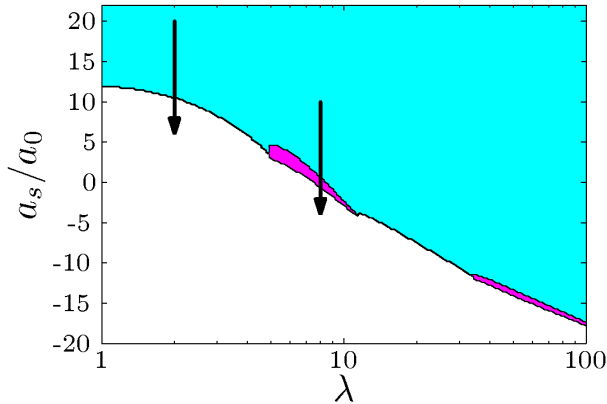}
\caption{\label{fig:stabclose}  (color online).  Stability diagram for $N=10^4$ $^{52}\mathrm{Cr}$ atoms.  The white region is dynamically unstable, while the darker regions are stable.  The pink (darker) islands are where biconcave structure is found in the ground state of the condensate.  The mean trap frequency is $\bar{\omega} = 2\pi \times 700$ Hz for all aspect ratios $\lambda$.  The arrows illustrate the initial and final values of scattering length in the experiment proposed in the text.}
\end{figure}

\subsection{More realistic experiment}

Figure~\ref{fig:phase} illustrates the kind of clean angular distribution that might be expected in the ideal experiment, where an infinitesimal change in scattering length is possible, and where only a single roton mode is excited.  Thus far, neither of these circumstances is true in the $^{52}\mathrm{Cr}$ experiments.  Whereas Figure~\ref{fig:ex8} shows the difference in scattering length at which modes with different $m$ quantum numbers develop imaginary energies to be a fraction of a Bohr, experimental uncertainties in the Feshbach-tuned scattering length of $^{52}\mathrm{Cr}$ are $\pm 2 a_0$~\cite{KochNature08a}.  Additonally, imaging of a $^{52}\mathrm{Cr}$ DBEC was done after a time of free-expansion in this experiment, not in-trap as is described in the scenario above.  We propose, with slight modification, an experiment similar to the one described by~\cite{KochNature08a} that presents us with the possibility of observing angular structure in the collapse and expansion of a DBEC.

Instead of making a very small jump in scattering length across the stability threshold, consider making a jump of $\Delta a_s = -14 a_0$.  For a $^{52}\mathrm{Cr}$ DBEC with $N=10^4$ atoms, we numerically prepare, for $\bar{\omega} = 2\pi \times 700$ Hz, a condensate in a trap with $\lambda = 2$ and scattering length $a_s = 20 a_0$ and a condensate in a trap with $\lambda = 8$ and scattering length $a_s = 10 a_0$, where both scattering lengths are about $10 a_0$ above $a_\mathrm{crit}$ for their respective aspect ratios.  We then ramp the scattering length from its initial value to its final value  over a time period of $8$ ms.  These scattering length ramps are illustrated by the arrows in Figure~\ref{fig:stabclose}.  Although an $8$ ms ramp time is not sufficiently slow to make the change completely adiabatic (the characteristic trap period is $2\pi / \bar{\omega} = 1.4 \; \mathrm{ms}$), it is sufficiently slow to allow a biconcave shape to form during the ramp.   Once this ramp has been made, we hold the collapsing condensate in the trap for $t_\mathrm{hold} = 2$ ms and then turn off the trap to let the collapsed condensate propagate in free space.  In an actual experiment, the expanded cloud could then be imaged to determine its density profile after expansion.  

To ensure that we accurately simulate an experimental scenerio and to break the cylindrical symmetry of the condensate in a physically consistent way, we seed the condensate prior to the time evolution with numerical noise.  Although the non-projected GPE can only model the condensate dynamics at zero temperature, an experiment will unavoidably have a very small but finite temperature present in the gas.  This means that the condensate fraction will not be exactly unity, but something slightly less.  To account for this, we correct our condensate by adding excited modes (quasiparticles) with weights determined by the Bose-Einstein distribution~\cite{Gardiner99},
\begin{equation}
\label{BED}
n_j = \left[e^{\frac{\omega_j - \mu }{ k_B T}} - 1\right]^{-1}
\end{equation}
where $n_j$ is the number of particles occupying the quasiparticle state with energy $\omega_j$, $T$ is the temperature of the Bose gas, $\mu$ is the chemical potential of the condensate and $k_B$ is the Boltzmann constant.  

Using the quasiparticle spectrum given by solving the BdG equations and a temperature of $T = 100$ nK, we then perturb our initial condensate by
\begin{eqnarray}
\label{pert}
\psi(\rr) \rightarrow \psi(\rho,z) + \sum_j \sqrt{\frac{n_j}{N}}\; e^{2\pi i \alpha_j} \left[u_{m,j}(\rho,z)e^{i m \varphi} \right. \nonumber \\
\left. + v^\star_{m,j}(\rho,z) e^{-i m \varphi} \right],
\end{eqnarray}
where $\{ \alpha_j\}$ are random numbers betwen $0$ and $1$, $n_j$ is given by Eq.~(\ref{BED}) and $u_{m,j}(\rho,z)$ and $v^\star_{m,j}(\rho,z)$ are BdG modes with quantum number $m$ and energy $\omega_j$.  Also, we include the factor $\sqrt{1/N}$ in the weighting because the condensate wavefunction $\psi(\rr)$ is normalized to unity instead of being normalized to $N$.  We impose a cutoff on the sum in Eq.~(\ref{pert}) of $\omega_j < 2k_B T$, where $T=100$ nK, simplifying the problem by ignoring higher energy modes that contribute little to the thermal excitations of the system.  Indeed, setting $T=100$ nK is an experimentally accessible temperature~\cite{Griesmaier05a}.

Additionally, because the condensate density becomes very large during the collapse process, a three-body loss term is required to accurately model the collapse and expansion dynamics~\cite{KochNature08a}.  The rate constant for three-body recombination was experimentally determined to be $L_3 = 2\times 10^{-40} \mathrm{m}^6 / \mathrm{s}$ for $^{52}\mathrm{Cr}$.  We account for this loss in our simulations by including the term $-i \hbar N(N-1) L_3 |\psi(\rr)|^4 / 2$ in the time-dependent GPE, given by
\begin{eqnarray}
\label{TDGPE}
i\hbar \frac{\partial \psi(\rr,t)}{\partial t} &=& \Big\{ -\frac{\hbar^2}{2M}\nabla^2 + U(\rr) \nonumber \\
&& + (N-1)\int d\rr^\prime V(\rr-\rr') |\psi(\rr^\prime,t)|^2 \nonumber \\
&& - N(N-1)\frac{i\hbar L_3}{2} |\psi(\rr,t)|^4 \Big\}\psi(\rr,t),
\end{eqnarray}
where $V(\rr-\rr^\prime)$ is given in Eq.~(\ref{V}).

\begin{figure*}
\includegraphics[width=2\columnwidth]{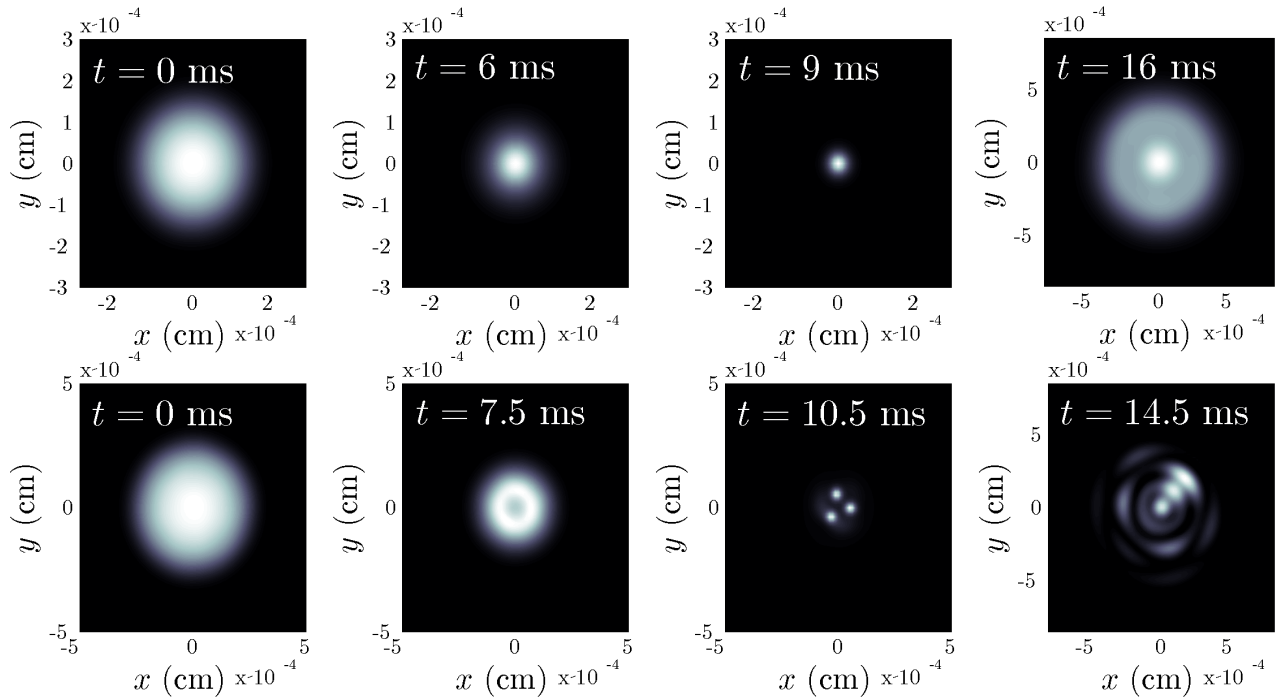}
\caption{\label{fig:expand}  Collapse dynamics of DBECs, both in harmonic traps with mean frequency $\bar{\omega} = 2\pi \times 700$ Hz, corresponding to the scattering length ramps illustrated in Figure~\ref{fig:stabclose}. (a)  A DBEC in a trap with aspect ratio $\lambda = 2$, at $t=0$ ms the condensate has $a_s = 20 a_0$, the scattering length is ramped down to $a_s = 6 a_0$ over $8$ ms, the collapsed condensate is held in the trap for $t_\mathrm{hold} = 2$ ms and then expanded until $t=16$ ms.  The collapse and expansion is purely radial.  (b) A DBEC in a trap with aspect ratio $\lambda = 8$, at $t=0$ ms the condensate has $a_s = 10 a_0$, the scattering length is ramped down to $a_s = -4 a_0$ over $8$ ms, the collapsed condensate is held in the trap for $t_\mathrm{hold} = 2$ ms and then expanded until $t=14.5$ ms.  The condensate becomes biconcave during the ramp in scattering length and thus collapses with angular structure, preserving an angular character during expansion.}
\end{figure*}

Figure~\ref{fig:expand} illustrates the numerical time evolution of these condensates through the collapse and expansion described above.  As before, these images represent density profiles in the $x$-$y$ plane, as viewed from the polarization axis.  
The top four frames illustrate the collapse and expansion of a condensate in a trap with aspect ratio $\lambda = 2$, in which there is no biconcave shape and in which, consequently, there should be no collapse to angular roton modes.  During its collapse, the condensate maintains its peak density in the center.  After the trap is removed and the gas is allowed to expand, its cylindrical symmetry is preserved.

By contrast, the lower four panels of Figure~\ref{fig:expand} illustrate a representative time evolution for a condensate in a trap wtih aspect ratio  $\lambda = 8$. In this case, by $7.5$ ms the condensate has established its biconcave structure.  When the condensate collapses, it does so into roton modes with angular nodal structure, leading to local collapse with angular nature.  After the trap is turned off and the condensate expands for $4.5$ ms, the angular structure remains in the density of the expanded cloud.  The collapse is clearly dominated by a roton with $m=3$ in this simulation.  However, because several angular modes are involved, the angular pattern no longer experiences pure $m=3$ angular symmetry. Moreover, each mode arrives with a random initial phase, meaning that there is a random assymmetry due to the interference between the unstable modes.  In the experiment this will imply non-repeatability of the observed density peaks from shot to shot.

Nevertheless, once the angular pattern is established, its vestiges remain in the expanded cloud.  In the final expanded picture, the clear break from cylindrical symmetry indicates that the decay modes have angular dependence, hence that the condensate went through a biconcave phase.

We note that a collapse and expansion experiment was done on a DBEC of $^{52}\mathrm{Cr}$~\cite{Metz09NJPhys}. However, it did not probe the parameter regime for biconcave structure formation.

The results of simulations that are very similar to the ones described above are presented in~\cite{Parker09}, where DBEC collapse is modeled in-trap and not through the expansion process and not including a three-body loss term in the simulation.  Ref.~\cite{Parker09} performs simulations of DBEC collapse for both adiabatic and non-adiabatic (instantaneous) jumps in scattering length, and find very interesting results regarding the presence of global and local collapse in the condensate dynamics.  For adiabatic collapse, where the change in scattering length is sufficiently slow to track the condensate across the roton softening in the BdG spectrum, they present a critical trap aspect ratio above which local collapse occurs.  We confirm these results, but point out that while local collapse is very interesting (and can be evidence for the presence of the roton in these systems), its manifestation in a DBEC is much richer than has been discussed in previous work.  A mapping of DBEC collapse via the experiment proposed above can determine not only whether collapse was global or local, but whether collapse was radial or angular and thus provide evidence for the underlying biconcave structure.

We point out that the experiment proposed above is just one of many experimental methods that would demonstrate the angular nature of DBEC collapse.  Certainly, taking data for a number of additional trap aspect ratios would assist in mapping out the regions where biconcavity exists.  Also, we expect that smaller and slower jumps in scattering length, which may be had with less uncertainty in the Feshbach-induced scattering length, would assist in understanding how and where collapse occurs.  Slower ramping of the scattering length allows the condensate to be tracked more adiabatically and thus allows for collapse to begin when only one BdG mode has a nonzero imaginary energy, making the mapping of the collapse much more clear.  Instantaneous or very fast jumps in scattering lengths across a biconcave region will miss this structure completely and thus result in a purely radial collapse.  For an angular collapse to occur, the biconcave structure must manifest itself in the condensate prior to collapse.

\section{Conclusion}
\label{sec:conclusion}

In conclusion, we have shown that in order to correctly map out the stability of a DBEC, a computational method that is sensitive to the local nature of DBEC collapse must be used.  Methods such as the gaussian \emph{ansatz} that are not sensitive to such phenomenon will incorrectly predict the stability of the system.  Also, we draw a connection between the BdG spectrum of a DBEC and the nature of the DBEC collapse.  Not only can the BdG quasiparticles predict where a DBEC will collapse in parameter space, they can also predict how a DBEC will collapse.  For DBECs without biconcave structure, this collapse is purely radial while for DBECs with biconcave structure, this collapse has angular structure.  Performing collapse and expansion experiments on a $^{52}\mathrm{Cr}$ DBEC can reveal this angular structure and thus provide an experimental method for mapping biconcave structure in DBECs.

\begin{acknowledgments}
The authors acknowledge the financial support of the U.S. Department of Energy and of the National Science Foundation, and thank the research group of Tilman Pfau for the experimental data.
\end{acknowledgments}

\appendix

\section{Calculation of the mean-field in reduced dimensions}
\label{app:meanfield}

We consider the calculation of the mean-field potential due to the dipole-dipole interaction in two different geometries, one with $\omega_\rho \rightarrow 0$ (a quasi-two dimensional (2D) geometry) and one with $\omega_z \rightarrow 0$ (a quasi-one dimensional (1D) geometry).  In the quasi-2D geometry, we assume that the condensate wavefunction depends only on $z$ and is homogeneous in the $\rho$-direction (or in the $x$- and $y$-directions) and in the quasi-1D geometry, we assume that the condensate wavefunction depends only on $\rho$ (or on $x$ and $y$) and is homogeneous in the $z$-direction.

We begin with the expression for the dipole-dipole interaction potential in momentum-space~\cite{Goral03}
\begin{equation}
\label{A1}
\tilde{V}_{dd}(\kk) = \frac{4\pi}{3} d^2 \left( 3\cos^2{\theta_\kk} - 1 \right),
\end{equation}
where $\theta_\kk$ is the angle between the direction of the dipole polarization ($\hat{z}$ or $\hat{k}_z$, for the DBEC we are considering) and the vector $\kk$.  Using this momentum-space representation, the coordinate-space mean-field potential due to the dipole-dipole interaction is given by the convolution of $\tilde{V}_{dd}(\kk)$ with the condensate density in momentum-space, $\tilde{n}(\kk)$,
\begin{equation}
\label{A2}
U_{dd}(\rr) = \mathcal{F}^{-1}\left[ \tilde{V}_{dd}(\kk) \tilde{n}(\kk) \right],
\end{equation}
where $\mathcal{F}^{-1}$ is the inverse Fourier Transform operator.  First, consider the quasi-2D geometry, in which the condensate density is homogeneous is $x$ and $y$.  The condensate density in momentum-space is then given by the Fourier Transform,
\begin{equation}
\label{A3}
\tilde{n}_\mathrm{2D}(\kk) = \mathcal{F}\left[ n_\mathrm{2D}(z) \right] = \tilde{n}_\mathrm{2D}(k_z)\delta(k_x)\delta(k_y).
\end{equation}
Substituting this result and Eq.~(\ref{A1}) into Eq.~(\ref{A2}) and writing $\cos^2{\theta_\kk} = k_z^2 / (k_x^2 + k_y^2 + k_z^2)$ gives an expression for the mean-field potential in the quasi-2D geometry,
\begin{equation}
\label{A4}
U^\mathrm{2D}_{dd}(\rr) = \mathcal{F}^{-1}\left[ \frac{4\pi}{3} d^2 \left( 3\frac{k_z^2}{k_x^2+k_y^2+k_z^2} - 1 \right) \tilde{n}_\mathrm{2D}(k_z)\delta(k_x)\delta(k_y) \right].
\end{equation}
The operation of the inverse Fourier Transform on this momentum-space function gives
\begin{equation}
\label{A5}
U^\mathrm{2D}_{dd}(\rr) = \frac{8\pi}{3}d^2 |\psi(z)|^2 = \frac{8\pi\hbar^2 a_{dd}}{M}|\psi(z)|^2,
\end{equation}
where $|\psi(z)|^2$ is the coordinate-space condensate density in the quasi-2D geometry.  We carry out the same calculation for the quasi-1D geometry, where the condensate density in momentum-space is given by
\begin{equation}
\label{A6}
\tilde{n}_\mathrm{1D}(\kk) = \mathcal{F}\left[ n_\mathrm{1D}(x,y)\right] = \tilde{n}_\mathrm{1D}(k_x,k_y) \delta(k_z).
\end{equation}
Substituting this function into Eq.~(\ref{A2}) gives
\begin{equation}
\label{A7}
U_{dd}^\mathrm{1D}(\rr) = -\frac{4\pi}{3} d^2 |\psi(\rho)|^2 = -\frac{4\pi \hbar^2 a_{dd}}{M} |\psi(\rho)|^2,
\end{equation}
where $|\psi(\rho)|^2$ is the coordinate-space condensate density in the quasi-1D geometry, written in terms of $\rho$ instead of $x$ and $y$.

\end{document}